\documentstyle[sprocl]{article}

\def\be{\begin{equation}}
\def\ee{\end{equation}}
\def\bea{\begin{eqnarray}}
\def\eea{\end{eqnarray}}

\begin{document}

\title{$p$--BRANE and $D$--BRANE ACTIONS}

\author{ E. BERGSHOEFF}

\address{Institute for Theoretical Physics, University of Groningen,\\
Nijenborgh 4, 9747 AG Groningen, The Netherlands}


\maketitle\abstracts{
We consider the actions for ten--dimensional $p$--branes and $D$--branes 
in an arbitrary curved background and discuss some of their properties.
We comment on how the $SL(2,R)$ duality symmetry acts
on the five--brane actions.}

\section{Introduction}

The original classification of super $p$--branes was based on the assumption
that the embedding coordinates, in a physical gauge, form worldvolume
scalar multiplets \cite{Ac1}. A classification of such scalar multiplets 
with $T$ (transverse) scalar degrees of freedom in $p+1$
dimensions leads to the following table:

\begin{table}[h]
\caption{Scalar multiplets with $T$ scalars in 
$p+1$ dimensions.\label{tab:sc}}
\vspace{0.4cm}
\begin{center}
\begin{tabular}{|c|c|c|c|c|}
\hline
$p+1$&$T$&$T$&$T$&$T$
\\ \hline
1&1&2&4&8\\
2&1&2&4&8\\
3&1&2&4&8\\
4&&2&4&\\
5&&&4&\\
6&&&4&\\ 
\hline
\end{tabular}
\end{center}
\end{table}

\noindent Each scalar multiplet corresponds to a $p$--brane in $d$ target
spacetime dimensions with 

\begin{equation}
d= (p+1) + T\, .
\end{equation}
Here the target space has been divided into
$p+1$ worldvolume and $T$ transverse directions.
The Table describes 16 $p$--branes with $p\le 5$ and $d\le 11$.

The corresponding $p$--brane actions consist of a kinetic term and
a Wess-Zumino (WZ) term. The kinetic term is given by

\begin{equation}
S_{\rm kin}^{(p)} = \int d^{p+1} \xi \ \sqrt {|g|}\, ,
\end{equation}
where $g$ is the determinant of the induced metric. The WZ term
is given by the pull-back of a $(p+1)$--form potential.
For the
case of interest (10 dimensions) these are a two--index
Neveu-Schwarz/Neveu-Schwarz (NS/NS) tensor $B^{(1)}$ (Type I,
IIA or IIB one--brane) 
and the dual six-index tensor ${\tilde B}^{(1)}_{\rm I}$ (Type I
five--brane):

\begin{eqnarray}
S_{\rm WZ}^{(1)} &=& \int d^{2}\xi\  B^{(1)}\, ,\label{WZ1}\\
S_{\rm WZ}^{(5)} &=&  \int d^{6}\xi\ {\tilde B}^{(1)}_{\rm I}\, ,
\label{WZ5}
\end{eqnarray}
where $d {\tilde B}_{\rm I}^{(1)} = {}^* d B^{(1)}$. Note that the action 
(\ref{WZ1}) is the same independent of whether the one--brane is
propagating in a $N=1$, IIA or IIB supergravity background.

\section{$D$--Branes}

Recently, a new class of extended objects has been introduced, called
Dirichlet branes or $Dp$--branes \cite{Pol1}.
They are described by embedding coordinates that, in a
physical gauge, form worldvolume vector multiplets.
A classification of all vector multiplets with $T$ (transverse)
scalars in $p+1$
dimensions is given by the table below.

\begin{table}[h]
\caption{Vector multiplets with $T$ scalar degrees of freedom in 
$p+1$ dimensions.\label{tab:ve}}
\vspace{0.4cm}
\begin{center}
\begin{tabular}{|c|c|c|c|c|}
\hline
$p+1$&$T$&$T$&$T$&$T$
\\ \hline
1&2&3&5&9\\
2&1&2&4&8\\
3&0&1&3&7\\
4&&0&2&6\\
5&&&1&5\\
6&&&0&4\\ 
7&&&&3\\
8&&&&2\\
9&&&&1\\
10&&&&0\\
\hline
\end{tabular}
\end{center}
\end{table}
In the case of ten dimensions, this leads to $Dp$--branes for
$0\le p \le 9$. The kinetic term of these $Dp$--branes is given 
by the following Born--Infeld type action:

\begin{equation}
S_{\rm kin}^{(Dp)} = \int d^{p+1}\xi\ e^{-\phi} \sqrt {|{\rm det}
(g_{ij} + {\cal F}_{ij})|}\, ,
\end{equation}
where $g_{ij}$ is the embedding metric and ${\cal F} = 2 d V -
B^{(1)}$ is the curvature of the worldvolume gauge field $V$.  
There is also a WZ term which describes the coupling of the 
Ramond--Ramond (RR) fields to the $Dp$--brane. In the case of the
$D1$--brane and $D5$--brane \cite{Gr1} these WZ terms are given by \footnote{
All terms are fixed by gauge invariance except for the last terms whose
coefficient can be determined via $T$--duality \cite{Be1}.}

\begin{eqnarray}
S_{\rm WZ}^{(D1)} &=& \int d^2\xi\  \bigl [
B^{(2)} + \ell {\cal F}\bigr ]\, ,\label{WZD1}\\
S_{\rm WZ}^{(D5)} &=& \int d^6\xi \  \bigl [
{\tilde B}_{{\rm IIB}}^{(2)} + 
{\textstyle\frac{1}{4}} \bigl (B^{(1)}\bigr )^2B^{(2)}
+ D{\cal F} \nonumber \\
&&{}\hskip .3truecm + {\textstyle\frac{3}{4}} B^{(1)}B^{(2)}{\cal F} +
{\textstyle\frac{3}{4}} B^{(2)}
{\cal F}^2 + \ell {\cal F}^3 \bigr ]\, ,\label{WZD5}
\end{eqnarray}
where ${\tilde B}_{{\rm IIB}}^{(i)}\ (i=1,2)$ is the IIB--dual of $B^{(i)}$:

\begin{equation}
d {\tilde B}_{{\rm IIB}}^{(i)} +\epsilon^{ij}
D d B^{(j)} - {\textstyle\frac{1}{4}}\epsilon^{ij}\epsilon^{kl}
B^{(j)}B^{(k)} d B^{(l)} = {}^* d B^{(i)}\, .
\end{equation}

The WZ terms of the Type IIB 1--brane in (\ref{WZ1}) and the D1--brane in 
(\ref{WZD1}), together with their kinetic terms,
are related to each other via a worldvolume Poincar\'e duality
that replaces the worldvolume vector $V$ by a constant $c$. In this proces
the background fields get replaced by their $S$-duals, e.g. $B^{(1)}$
gets replaced by $B^{(2)} - c B^{(1)}$, etc. Together with a shift symmetry
of the scalar $\ell$, this leads to a realization of the $SL(2,R)$
duality group on the Type IIB 1--brane actions (\ref{WZ1}), (\ref{WZD1})
\cite{Sc1}.

The situation differs for the WZ terms of the
5--brane (\ref{WZ5}) and the D5--brane (\ref{WZD5}).
The reason for this is that the 5--brane (\ref{WZ5})
is a Type I 5--brane with $4+4$
worldvolume degrees of freedom whereas the $D5$--brane is a Type
IIB 5--brane with $8+8$ degrees of freedom. 
To realize the $SL(2,R)$ duality symmetry on the 
$D5$--brane action, we need a second Type IIB
5--brane action. For this purpose we cannot consider the direct dimensional
reduction of the eleven--dimensional five--brane since this reduction
leads to a Type IIA five--brane that naturally couples to the Type IIA dual
potential ${\tilde B}_{{\rm IIA}}^{(1)}$:

\begin{equation}
d {\tilde B}_{{\rm IIA}}^{(1)} - {\textstyle\frac{105}{4}} C d C - 
7 A^{(1)} G({\tilde C}) = {}^* d B^{(1)}\, .
\end{equation}
Here $G({\tilde C})$ is the curvature of the dual 5--form potential
${\tilde C}$ defined in \cite{Be2}.

In analogy with the the case of the one--brane described above we
expect that under a
worldvolume Poincar\'e duality transformation the $D5$--brane gets converted 
into another Type IIB 5--brane where the vector field $V$ is replaced
by a 3--form gauge field $W$. Due to the presence of higher--order ${\cal F}$
terms in the WZ term (\ref{WZD5})
this duality transformation will be nonlinear and difficult to perform
explicitly. Perhaps a better approach, based on an analogy with a formulation
of the $D4$--brane action \cite{Be2}, is to work with a 1--form $V$ and 
a 3--form $W$ 
at the same time, and
to eliminate one or the other at the level of the field equations
via a curvature constraint. We find that the gauge transformations
and curvature for $W$ are given by

\begin{eqnarray}
\delta W &=& \rho +  3\Sigma^{(2)} d V \, ,\nonumber\\
{\cal H} &=& d W - D -{\textstyle\frac{3}{4}}B^{(1)}B^{(2)}
-{\textstyle\frac{3}{2}} B^{(2)}{\cal F}\, .
\end{eqnarray}
Our next task is to construct a self--dual action such that the
constraint ${\cal H} = e^{-\phi}{}\ ^* {\cal F}$ can be imposed consistently.
It seems that this is indeed possible. It would be interesting to
see how the $SL(2,R)$ symmetry works in the context of this selfdual
action.

\section*{Acknowledgments}
I would like to thank the organizers of the Second International
Sakharov Conference on Physics for their kind invitation to give this
talk. This work has been made possible by a fellowship
of the Royal Dutch Academy of Arts and Sciences (KNAW).

\end{document}